\documentstyle[aps,prl]{revtex}
\tighten
\begin{document}
\preprint{DAMTP R96/23,\,FERMILAB-Pub-96/265-A,\,WISC-MILW-96/21}
\draft

%%%%%% defines must be replaced before submitting electronically %%%%%%%%%%%%

\def\mathrelfun#1#2{\lower3.6pt\vbox{\baselineskip0pt\lineskip.9pt
  \ialign{$\mathsurround=0pt#1\hfil##\hfil$\crcr#2\crcr\sim\crcr}}}
\def\simlt{\mathrel{\mathpalette\mathrelfun <}}
 
%%%%%%%%%%%%%%%%%%%%%%%%%%%%%%%%%%%%%%%%%%%%%%%%%%%%%%%%%%%%%%%%%%%%

\title{Large Angular Scale CMB Anisotropy Induced by Cosmic Strings}

\author{B. Allen$^1$, R. R. Caldwell$^2$, E. P. S. Shellard$^2$,
	A. Stebbins$^3$, and S. Veeraraghavan$^4$}
\address{${}^1$Department of Physics, University of Wisconsin --
	Milwaukee, P.O. Box 413, Milwaukee, Wisconsin 53201, U.S.A.}
\address{${}^2$University of Cambridge, D.A.M.T.P.
	 Silver Street, Cambridge CB3 9EW, U.K.}
\address{${}^3$NASA/Fermilab Astrophysics Center,
	 P.O. Box 500, Batavia, Illinois 60510, U.S.A.}
\address{$^4$NASA/Goddard Space Flight Center, Greenbelt,
	Maryland 20771}

\twocolumn[

We simulate the anisotropy in the cosmic microwave background (CMB)
induced by cosmic strings. By numerically evolving a network of cosmic
strings we generate full-sky CMB temperature anisotropy maps. Based on
$192$ maps, we compute the anisotropy power spectrum for multipole
moments $\ell \le 20$.  By comparing with the observed temperature
anisotropy, we set the normalization for the cosmic string
mass-per-unit-length $\mu$, obtaining $G\mu/c^2=1.05 \,
{}^{+0.35}_{-0.20} \times10^{-6}$, which is consistent with all other
observational constraints on cosmic strings.  We demonstrate that the
anisotropy pattern is consistent with a Gaussian random field on large
angular scales.

\maketitle

\pacs{PACS numbers: 98.80.Cq, 11.27.+d, 98.70.Vc}
]

\narrowtext

%%%%%%%%%%%%%%%%%%%%%%%%%%%%%%%%%%%%%%%%%%%%%%%%%%%%%%%%%%%%%%%%%%%%

%%%%%%%%%%%%%%%%%%%%%%%%%%%%%%%%%%%%%%%%%%%%%%%%%%%%%%%%%%%%%%%%%%%%
% Introduction
%%%%%%%%%%%%%%%%%%%%%%%%%%%%%%%%%%%%%%%%%%%%%%%%%%%%%%%%%%%%%%%%%%%%

Cosmic strings are topological defects which may have formed in the
very early universe and may be responsible for the formation of large
scale structure observed in the Universe today \cite{StringReviews}.
In order to test the hypothesis that the inhomogeneities in our
universe were induced by cosmic strings one must compare observations
of our universe with the predictions of the cosmic string model. This
{\it Letter} presents results of detailed computations of the large
angular scale cosmic microwave background (CMB) anisotropies induced by
cosmic strings \cite{OurPastResults}.  These predictions are compared to 
the large scale anisotropies observed by the COsmic Background
Explorer (COBE) satellite.  Because the predicted temperature
perturbations are proportional to the dimensionless quantity $G
\mu/c^2$ where $G$ is Newton's constant and $c$ the speed of light, one
may constrain the value of $\mu$, the mass per unit length of the
cosmic strings.  We believe that our estimate of $\mu$ is the most
accurate and reliable to date.

%%%%%%%%%%%%%%%%%%%%%%%%%%%%%%%%%%%%%%%%%%%%%%%%%%%%%%%%%%%%%%%%%%%%
% Procedure
%%%%%%%%%%%%%%%%%%%%%%%%%%%%%%%%%%%%%%%%%%%%%%%%%%%%%%%%%%%%%%%%%%%%

Our methodology for computing the large angle anisotropy is to simulate
the evolution of random realizations of a cosmic string network
\cite{NumSim}. From these network simulations we construct the
temperature anisotropy pattern seen by various observers within the
simulation volume.  We have evolved the strings from a redshift $z=100$
to the present, in a cubical box whose side length is twice the Hubble
radius at the end of the simulation.  This large box assures that the
anisotropy pattern is unaffected by the finite simulation volume.

In order to obtain the large dynamic range required for these
simulations we have used a new technique whereby the number of segments
used to represent the string network decreases as the simulation
proceeds.  We have conducted tests of this method by comparing smaller
simulations, with and without decreasing the number of segments: the
average long string energy density is unaffected; the distribution of
coherent velocities (the string velocity averaged over a particular
length scale) is preserved down to scales smaller than $1/100$ of the
horizon radius; the effective mass-per-unit-length of string (the
energy in string averaged over a particular length scale) is preserved
down to scales smaller than $1/100$ of the horizon radius. The decrease
in the number of segments was regulated so that on the angular scales
of interest the simulation provided a good representation of the cosmic
string network.  Here we are interested in comparing with data from the
COBE Differential Microwave Radiometer (DMR) which measures the
anisotropy convolved with an approximately $7^\circ$ FWHM beam
\cite{DMR}.  Our contact with this data is through COBE's predicted
correlation function at the $10^\circ$ angular scale.

The temperature patterns are computed using a discretized version of the
integral equation
\begin{equation}
{\Delta T\over T}(\hat{n},x_{\rm obs})=
\int d^4x\,G^{\mu\nu}(\hat{n},x_{\rm obs},x)\Theta_{\mu\nu}(x)
\end{equation}
where $\Theta_{\mu\nu}$ is the stress-energy tensor of the string at
4-position $x$, $x_{\rm obs}$ is the 4-position of the observer, and
$\hat{n}$ indicates the direction of the arrival of photons on the
celestial sphere.  Our results apply only if the present spatial
curvature and cosmological constant are small since we use a Green's
function $G^{\mu\nu}$ appropriate to a matter-dominated
Einstein-deSitter cosmology \cite{Green}.  While the universe is not
entirely matter-dominated at redshifts close to recombination ($z_{\rm
rec} \approx 1100$), the approximation is justified because we show
below that the large-scale anisotropies are primarily produced at
$z\simlt100$.  Hydrodynamical effects at recombination, which only
slightly affect the large angle anisotropy, are not included here.
Instead we assume that that the photons are comoving with the dark
matter at recombination.  We have used the ansatz of local compensation
\cite{VeeraraghavanStebbins} as the initial condition for the
perturbations of the matter distribution.

%%%%%%%%%%%%%%%%%%%%%%%%%%%%%%%%%%%%%%%%%%%%%%%%%%%%%%%%%%%%%%%%%%%%
% Power Spectrum
%%%%%%%%%%%%%%%%%%%%%%%%%%%%%%%%%%%%%%%%%%%%%%%%%%%%%%%%%%%%%%%%%%%%

We have generated three independent realizations of the cosmic string
network.  For each realization we have computed the fractional CMB
temperature perturbation, ${\Delta T \over T} (\hat{n},x_{\rm obs})$,
in 6144 pixel directions, $\hat{n}$, on the celestial sphere of 64
observers distributed uniformly throughout each simulation box.  This
computational scheme gives ${\Delta T \over T}$ smoothed on about the
size of the pixels which are $~3.5^\circ$ or smaller
\cite{PapersToWrite}.  One of these temperature maps is shown in figure
\ref{PrettyPicture}. This temperature map has been smoothed with a
$10^\circ$ FWHM beam to permit direct comparison with the published
COBE sky maps ($\rm
http://www.gsfc.nasa.gov/astro/cobe/dmr\_image.html $).  Note that
these published maps have structure on angular scales smaller than
$10^\circ$; this is receiver noise.

Each temperature map may be
expressed in terms of the scalar spherical harmonics $Y_{\ell m}$ on
the sphere:
\begin{equation}
{\Delta T \over T}(\hat{n},x_{\rm obs}) = \sum_{\ell=0}^{\infty} \sum_{m=-\ell}^\ell
a_{\ell m}(x_{\rm obs}) Y_{\ell m}(\hat{n}).
\end{equation}
In this {\it Letter} we only consider coefficients $a_{\ell m}$ with
$\ell\le20$. For $\ell>20$ the error due to finite pixel size and
gridding effects is larger than $10\%$. In reference
\cite{PapersToWrite} we consider how to correct the large $\ell$
harmonics for discretization effects.

For each map we construct the multipole moments
\begin{equation}
\widehat{C}_\ell={1\over2\ell+1}\sum_{m=-\ell}^\ell |a_{\ell m}(x_{\rm obs})|^2.
\end{equation}
The monopole and dipole moments ($\ell=0,1$) are discarded because they
contain no useful information.  The mean and variance of
$\ell(\ell+1)\widehat{C}_\ell$ for $2 \le \ell \le 20$ is plotted in
figure \ref{fig_llcl_sim123} in units of $(G \mu/c^2)^2$.  We see that
the mean spectrum is roughly flat, with
$\ell(\ell+1)\widehat{C}_\ell\sim 350 (G\mu/c^2)^2$ for $\ell\lesssim
20$.

The standard estimator for the ensemble-averaged correlation function
$\widehat C(\gamma)$ of ${\Delta T \over T}$ at angle $\gamma$ may be
expressed in terms of the $\widehat C_\ell$. We smooth the temperature
pattern first with the average DMR beam model window function $G_\ell$
(tabulated values are given in \cite{WrightsBeam}) which is
approximately a $7^\circ$ beam, and second with a $7^\circ$ FWHM
Gaussian window function $W_\ell(7^\circ)$, for an effective smoothing
on angular scales $\theta_{\scriptscriptstyle \rm smooth} = 10^\circ$.
\begin{eqnarray}
\widehat{C}(\gamma,10^\circ)
	&=&\sum_{\ell=2}^{20}{2\ell+1\over4\pi}\widehat{C}_\ell\,
 	|G_\ell|^2 |W_\ell(7^\circ) |^2 {\rm P}_\ell(\cos \gamma) 
							\cr\cr
W_\ell(\theta)
	&=& {\rm exp} \Bigl( -{\ell(\ell+1)\over \ln{2}}
	 \sin^2{\theta\over 4} \Bigr)
\end{eqnarray}
Here the ${\rm P}_\ell$ are Legendre polynomials. By neglecting
$\ell>20$ we would underestimate this sum for
$\theta_{\scriptscriptstyle\rm smooth}<10^\circ$. We compute the
correlation $\widehat{C}(0^\circ,10^\circ)$ at the angle
$\gamma=0^\circ$ from our cosmic string simulations and compare it with
the same quantity as estimated by the COBE team from the COBE data.

We expect that most of the anisotropy on a particular angular scale is
seeded by the strings near the time when the projection of the
coherence length of the string network subtends that angle on the
celestial sphere \cite{BBS,BSB,PerivolaropoulosModel,Hara93}. Since the
coherence length of the string network grows with time, we expect
anisotropies on small angular scales to be seeded at early times, and
the large angle anisotropies to be seeded at late times. We have
constructed temperature maps which include only a part of the temporal
evolution of the string network, from a redshift $z$ to the present.
By considering the convergence of $\widehat{C}(0^\circ, 10^\circ)$ as
$z$ increases we may determine what redshift range is required to
accurately determine $\widehat{C}(0^\circ, 10^\circ)$.  In figure
\ref{fig_c0_sim123} we plot the average value of
$\widehat{C}(0^\circ,10^\circ)$ in units of $(G\mu/c^2)^2$ for 64
observers in each of three string simulations as a function of $z$.  We
see that $\widehat{C}(0^\circ, 10^\circ)$ receives its dominant
contribution within a redshift $z\sim 20$, although there is continued
growth through $z\sim 100$. Two independent models of the temperature
correlation function indicate that by neglecting the contribution of
cosmic strings in the redshift range $100<z<z_{\rm rec}$ we
underestimate $\widehat{C}(0^\circ,10^\circ)$. The underestimate in the
smoothed autocorrelation function is $18\%$ according to the
semi-analytic model of Bennett {\it et al} \cite{BSB}, and $25\%$
according to the analytic model of Perivolaropoulos
\cite{PerivolaropoulosModel}. We use the latter, more conservative
model to extrapolate our results, valid to $z=100$, out to the redshift
of recombination.

We may normalize the cosmic string mass per unit length, $\mu$, by
matching estimates of $\widehat{C}(0^\circ,10^\circ)$ from COBE-DMR
with our predictions.  The COBE-DMR four year sky maps yield
$\widehat{C}(0^\circ,10^\circ)=(29\pm1\,\mu{\rm K}/T)^2$\cite{COBE4RMS}
with mean temperature $T=2.728\pm0.002\,$K \cite{CMBRtemperature}.  Our
simulations indicate that at $z=100$, $\widehat{C}(0^\circ,10^\circ)=
82 \pm 19 (G\mu/c^2)^2$ where the $\pm 19$ gives the cosmic variance
between the different observers in all three of our simulations.  The
analytic model \cite{PerivolaropoulosModel} predicts that at $z_{\rm
rec}$, $\widehat{C}(0^\circ,10^\circ) = 103 \pm 24 \pm 20(G\mu/c^2)^2$,
where the $\pm 24$ is the new cosmic variance.  The $\pm 20$ is a
conservative systematic error composed of a $\sim 10 \%$ uncertainty
due to the simulation technique of reducing the number of string
segments, $\sim 7 \%$ due to the difference in the two models used for
the extrapolation out to $z_{rec}$, and $\sim 5 \%$ due to the
discretization of the celestial sphere. (These errors will be discussed
in more detail in \cite{PapersToWrite}.) Note that our result is not
strongly dependent on the extrapolation out to $z_{\rm rec}$, which
makes only a small correction, lowering the normalization of $\mu$ by
$10 \%$, in comparison to the quoted uncertainties.  Hence, adding
these errors linearly, normalization to COBE yields
\begin{equation}
G\mu/c^2=1.05 \,{}^{+0.35}_{-0.20} \times10^{-6} 
\label{Normalization}
\end{equation}
for the cosmic string mass per unit length.

%%%%%%%%%%%%%%%%%%%%%%%%%%%%%%%%%%%%%%%%%%%%%%%%%%%%%%%%%%%%%%%%%%%%
% Gaussianity
%%%%%%%%%%%%%%%%%%%%%%%%%%%%%%%%%%%%%%%%%%%%%%%%%%%%%%%%%%%%%%%%%%%%

Because the spatial distribution of the cosmic string network is not
described by Gaussian random variables, we expect that the anisotropy
pattern generated in the cosmic string scenario will be non-Gaussian at
some level.  At very small angular scales, the sharp temperature
discontinuities across strings guarantee non-Gaussian features.  One
might hope to find a non-Gaussian signature to distinguish cosmic
string models from inflationary models, for which the anisotropy
patterns are expected to be very Gaussian.  On the large DMR angular
scales we are studying, however, we will see that many different
strings contribute significantly to each resolution element of the
temperature pattern, so the conditions for the central limit theorem
are very well satisfied and the temperature pattern is very close to
Gaussian.

We have looked for non-Gaussianity in the distribution function of the
temperature anisotropy after smoothing our maps with the average DMR
beam \cite{WrightsBeam} model window function, an approximately
$7^\circ$ Gaussian beam. The distribution function after combining all
of our maps is shown in figure \ref{fig_pixdist_total}. We see that the
distribution, on angular scales accessible to DMR, is very close to a
Normal distribution. The pixel temperature distribution for any such
single observer will not appear as smooth, just as for a limited sample
drawn from a true Normal distribution.

It has been suggested that the distribution of temperature differences
is a better test of non-Gaussianity than the temperature distribution
\cite{GradientIdea,PiInSky}. In figure \ref{fig_gradient_total} we plot
the distribution of the differences in temperature of nearby pixels.
We see that for temperature differences on angular separations greater
than the COBE DMR $7^\circ$ angular resolution scale, the distribution
of temperature differences is again very close to a Normal
distribution.  With finer angular resolution than that probed by DMR,
as may be possible with the MAPS or COBRAS/SAMBAS detectors, the
inherently non-Gaussian character of the temperature anisotropy due to
cosmic strings may be observed. We shall further examine this and other
aspects of the map statistics in reference \cite{PapersToWrite}.

%%%%%%%%%%%%%%%%%%%%%%%%%%%%%%%%%%%%%%%%%%%%%%%%%%%%%%%%%%%%%%%%%%%%
% Summary
%%%%%%%%%%%%%%%%%%%%%%%%%%%%%%%%%%%%%%%%%%%%%%%%%%%%%%%%%%%%%%%%%%%%

In this {\it Letter} we have presented the first computation of large
angle CMB anisotropy from cosmic strings which has included all of the
relevant physics.  Our normalization of $\mu$ (\ref{Normalization}) is
consistent with most previous work. Existing studies of the expected
large-scale CMB anisotropies find $G\mu/c^2 = 1.5 (\pm 0.5) \times
10^{-6}$ \cite{BSB} and $1.7 (\pm 0.7) \times 10^{-6}$
\cite{PerivolaropoulosModel} (also see \cite{Hara93}). Note that our
results do not appear consistent with Coulson {\it et al}
\cite{PiInSky}, who obtain the higher normalization $G\mu/c^2 = 2
\times 10^{-6}$ (they quote no uncertainties). They use the
lattice-based Smith-Vilenkin evolution algorithm in Minkowski
space-time to produce $18$ realizations of a $30^\circ $ square patch
temperature field. Our results should be more accurate because:  we
simulate the string motion in an expanding universe; our simulation
allows smooth variation of quantities such as the string's velocity;
our $192$ full-sky realizations are comparable to over $8800$
$30^\circ$ patches and thus have better statistics.  Our normalization
of $\mu$ is also compatible with other observational constraints on
cosmic strings. The bound on gravitational radiation due to pulsar
timing residuals and primordial nucleosynthesis gives $G \mu/c^2 < 5.4
(\pm 1.1) \times 10^{-6}$ \cite{GravityWaveBound}.  The bound due to
cosmic rays emitted by evaporating primordial black holes formed from
collapsed cosmic string loops gives $G\mu/c^2 < 3.1 (\pm 0.7) \times
10^{-6}$ \cite{CaldwellCasper}.  CMB fluctuations on angular scales
below $10$ arc-seconds gives $G\mu/c^2 < 2 \times 10^{-6}$ at the $95
\%$ confidence level, assuming no reionization \cite{Hindmarsh}.  The
normalization presented in this {\it Letter} should allow a more direct
confrontation of the cosmic string model with observations of large
scale density inhomogeneities.

%%%%%%%%%%%%%%%%%%%%%%%%%%%%%%%%%%%%%%%%%%%%%%%%%%%%%%%%%%%%%%%%%%%%
% Acknowledgements
%%%%%%%%%%%%%%%%%%%%%%%%%%%%%%%%%%%%%%%%%%%%%%%%%%%%%%%%%%%%%%%%%%%%
\acknowledgements
\vspace{-0.1in}
We thank the LASP at NASA/Goddard Space Flight Center for use of
computing resources. The work of BA was supported by NSF grants
PHY91-05935 and PHY95-07740. The work of RRC and EPSS was supported by
PPARC through grant GR/H71550. The work of AS was supported by the DOE
and by NASA under grant NAGW-2788. The work of SV was supported by the
NRC at Goddard.

%%%%%%%%%%%%%%%%%%%%%%%%%%%%%%%%%%%%%%%%%%%%%%%%%%%%%%%%%%%%%%%%%%%%
% References
%%%%%%%%%%%%%%%%%%%%%%%%%%%%%%%%%%%%%%%%%%%%%%%%%%%%%%%%%%%%%%%%%%%%

%%%%%%%%%%%%%%%%%%%%%%%%%%%%%%%%%%%%%%%%%%%%%%%%%%%%%%%%%%%%%%%%%%%%
% Figures
%%%%%%%%%%%%%%%%%%%%%%%%%%%%%%%%%%%%%%%%%%%%%%%%%%%%%%%%%%%%%%%%%%%%
\vspace{0.1in}
\centerline{\bf FIGURES}

\begin{figure}
\caption{
The temperature anisotropy map for one of 192 realizations of the
cosmic string induced CMB anisotropy patterns which we have generated
for this {\it Letter}. This is an equal area, all-sky map using a
Hammer-Aitoff projection.  The map has been smoothed with a $10^\circ$
FWHM Gaussian beam.}
\label{PrettyPicture}
\end{figure}

\vspace{-0.2in}
\begin{figure}
\caption{
Plotted is the $\ell(\ell+1)\widehat{C}_\ell$ angular power spectrum
for the anisotropies from our realizations. The
central dot gives the mean for all 192 observers while the symmetric
error bars give the rms variation between different observers.}
\label{fig_llcl_sim123}
\end{figure}

\vspace{-0.2in}
\begin{figure}
\caption{
Plotted is the rms anisotropy after smoothing with an effective $10^\circ$
smoothing beam versus the maximal redshift, $z$, for which
strings have been included. The different curves give the mean value of
rms anisotropy averaged over the 64 observers in each of the three
cosmic string simulations. Local compensation has been applied at each
starting time.}
\label{fig_c0_sim123}
\end{figure}

\vspace{-0.2in}
\begin{figure}
\caption{
Plotted is the frequency of temperature anisotropy for all pixels in
each of our 192 maps after smoothing with a model of the average DMR
beam, an approximately $7^\circ$ Gaussian beam.  The solid line gives a
Normal distribution with zero mean and variance determined by the pixel
distribution amplitude.}
\label{fig_pixdist_total}
\end{figure}

\vspace{-0.2in}
\begin{figure}
\caption{
Plotted is the frequency of temperature differences for all pairs of
pixels at a given angular separation in each of our 192 maps with no
smoothing. The lines show a Normal distribution with variance
determined by the pixel difference distribution amplitudes.  The three
pairs of curves with increasing variance are for separations spanning the
ranges
$10^\circ\pm 1^\circ$, $20^\circ\pm 1^\circ$, and $40^\circ\pm
1^\circ$.}
\label{fig_gradient_total}
\end{figure}


\begin{references}
\vskip -0.4in
\centerline{\bf REFERENCES}

\bibitem{StringReviews}
A.~Vilenkin,  Phys. Rep. {\bf 121}, 263 (1985); A.~Vilenkin and
E.~P.~S.~Shellard, {\it Cosmic Strings and other Topological Defects},
(Cambridge University Press, Cambridge, England, 1994); M. Hindmarsh
and T.W.B. Kibble, Rep. Prog. Phys. {\bf 58}, 47 (1995).


\bibitem{OurPastResults}
B.~Allen {\it et al.}, in {\it CMB Anisotropies Two Years After COBE}, ed.
L.~Krauss (World Scientific, New York, 1993).

\bibitem{NumSim}
B.~Allen    and E.~P.~S.~Shellard, Phys. Rev. Lett. {\bf  64},  119 (1990).

\bibitem{DMR}
G.~Smoot {\it et al.}, Ap.~J {\bf 360}, 685 (1990).

\bibitem{Green}
S.~Veeraraghavan and A.~Stebbins, in preparation (1996).

\bibitem{VeeraraghavanStebbins}
S.~Veeraraghavan and A.~Stebbins, Ap.~J. {\bf 365}, 37 (1990).

\bibitem{PapersToWrite}
B.~Allen {\it et al.}, in preparation (1996).

\bibitem{WrightsBeam}
E.~L.~Wright {\it et al.}, Ap.~J {\bf 420}, 1 (1994).

\bibitem{BBS}
F.~Bouchet, D.~Bennett, and A.~Stebbins, Nature {\bf 355}, 410 (1988).

\bibitem{BSB}
D.~Bennett, A.~Stebbins, and F.~Bouchet, Ap.~J.~Lett. {\bf 399}, L5 (1992);

\bibitem{PerivolaropoulosModel}
L.~Perivolaropoulos, Phys. Lett. {\bf B298}, 305 (1993).

\bibitem{Hara93}
T.~Hara, P.~Mahonen, and S.~Miyoshi, Ap.~J. {\bf 414}, 421 (1993).

\bibitem{COBE4RMS}
A.~Banday  {\it et al.} COBE Preprint 96-04 astro-ph/9601065 (1996).

\bibitem{CMBRtemperature}
C.~Bennett {\it et al.} COBE Preprint 96-01 astro-ph/9601067 (1996).

\bibitem{GradientIdea}
R.~Moessner, L.~Perivolaropoulos, and R.~Brandenberger, Ap.~J. {\bf 425}, 365
(1994);
P.~Graham, N.~Turok, P.~Lubin, and J.~Schuster, Ap.~J. {\bf 449}, 404 (1995).

\bibitem{PiInSky}
D.~ Coulson, P.~Ferreira, P.~Graham, and N.~Turok, Nature {\bf 368},
27 (1994).

\bibitem{GravityWaveBound}
R.~R.~Caldwell, R.~A.~Battye, and E.~P.~S.~Shellard, University of
Cambridge Preprint DAMTP-R96-21, astro-ph/9607130.


\bibitem{CaldwellCasper}
R.~R.~Caldwell and Paul Casper, Phys. Rev. {\bf D53}, 3002 (1996).

\bibitem{Hindmarsh}
Mark Hindmarsh, Ap. J. {\bf 431}, 534 (1994).

\end{references}
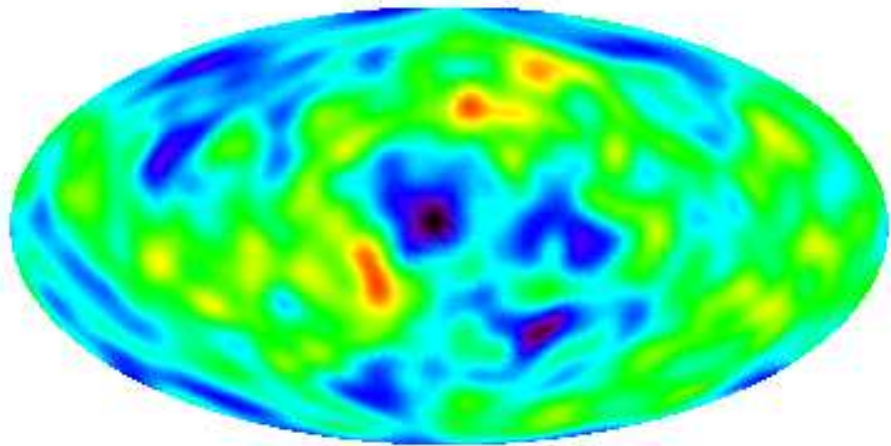
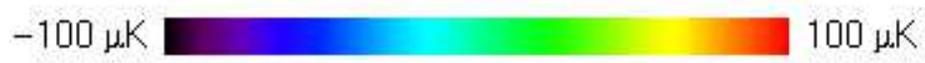
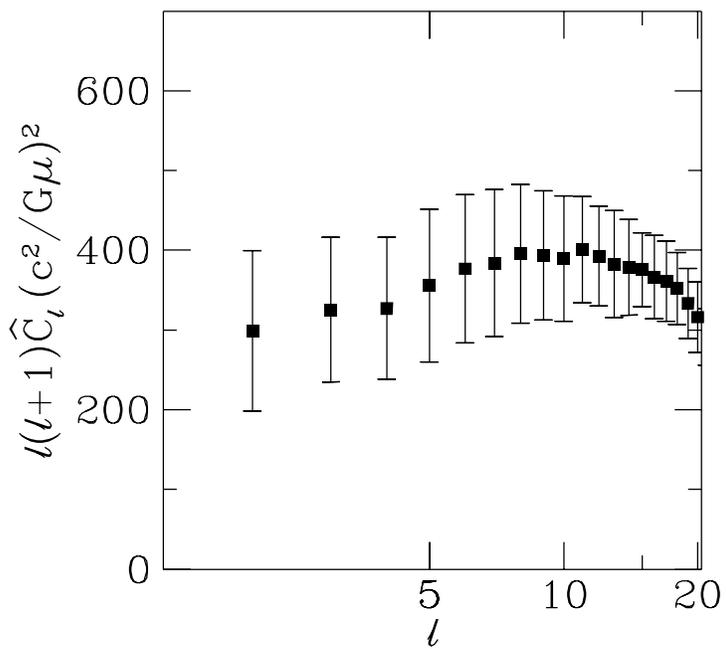
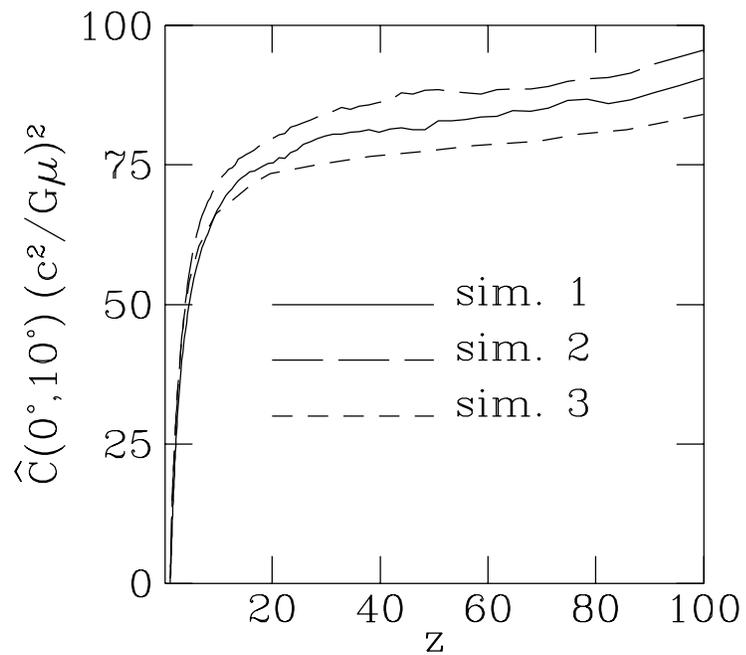
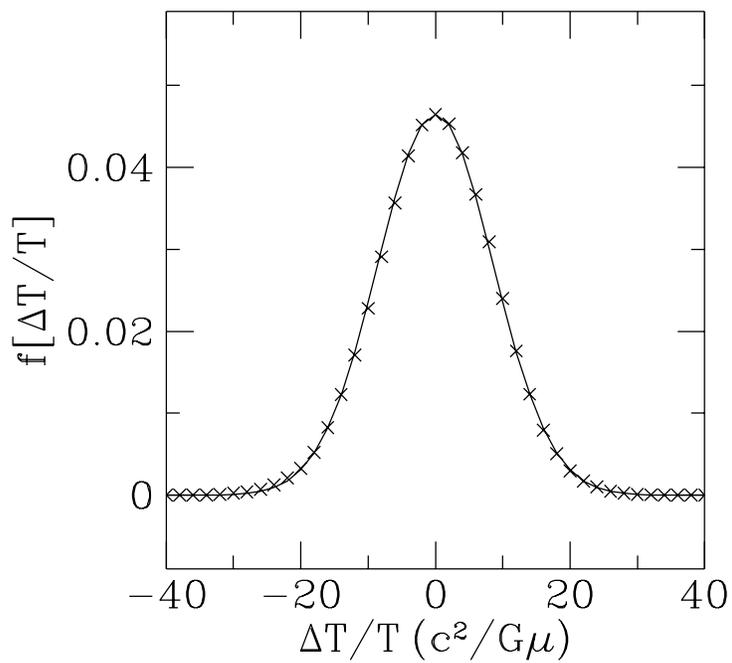
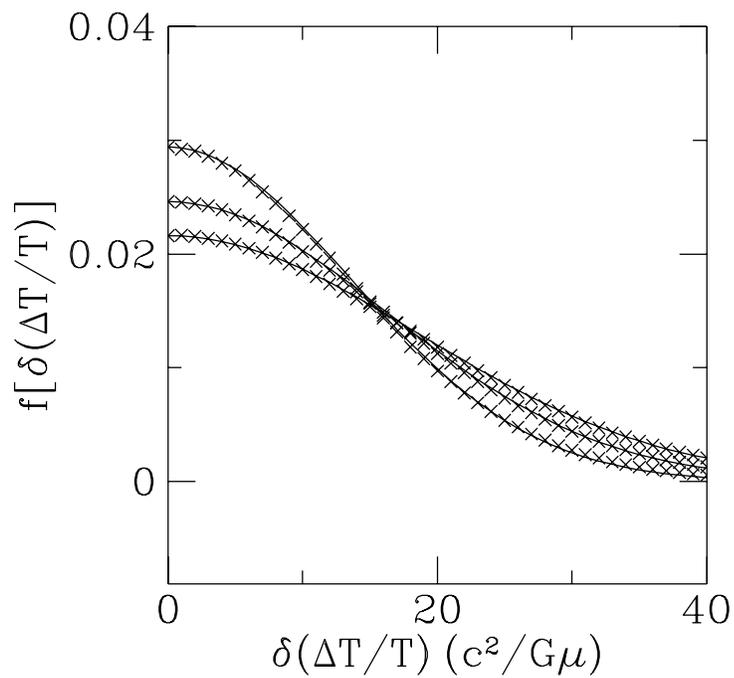
\end{document}